\documentclass[aps,reprint]{revtex4-1}
\usepackage[utf8]{inputenc}
\usepackage{empheq}
\usepackage{amsmath}
\usepackage{amsfonts}
\usepackage{amssymb,bm,txfonts}
\usepackage{mathtools}
\usepackage{stmaryrd}
\usepackage{dsfont}
\usepackage{graphicx}
\usepackage[bookmarksnumbered=true]{hyperref}
\usepackage{stackrel}
\usepackage[usenames]{color}
\usepackage{comment}
\usepackage{xcolor}

\usepackage{graphicx}
\usepackage{siunitx}
\usepackage{fancyhdr}

\def \ee{\end{equation}}
\def \be{\begin{equation}}
\def \eea{\end{eqnarray}}
\def \bea{\begin{eqnarray}}

\newcommand{\eqnref}[1]{Eqn.~(\ref{#1})}		% for equations with preceeding Eqn.
\newcommand{\eqnsref}[1]{Eqns.~(\ref{#1}}       % for several equations. Give \eqnsref{ref1}--\ref{ref2})
\newcommand{\cannex}{{\sc{Cannex}}}
\begin{document}
\title{Vacuum energy, the Casimir effect, and Newton's non-constant}
\author{Benjamin Koch}
\email{benjamin.koch@tuwien.ac.at}
\affiliation{Institut f\"ur Theoretische Physik,
 Technische Universit\"at Wien,
 Wiedner Hauptstrasse 8--10,
 A-1040 Vienna, Austria}
 \affiliation{ Instituto de F\'isica, Pontificia Universidad Cat\'olica de Chile, 
Casilla 306, Santiago, Chile}
\affiliation{Atominstitut, Technische Universit\"at Wien,  Stadionallee 2, A-1020 Vienna, Austria}
\author{Christian K\"ading}
\affiliation{Atominstitut, Technische Universit\"at Wien,  Stadionallee 2, A-1020 Vienna, Austria}
\author{Mario Pitschmann}
\affiliation{Atominstitut, Technische Universit\"at Wien,  Stadionallee 2, A-1020 Vienna, Austria}
\author{Ren\'e I. P. Sedmik}
\affiliation{Atominstitut, Technische Universit\"at Wien,  Stadionallee 2, A-1020 Vienna, Austria}
\begin{abstract}
We explore two hypotheses.
First,
the possibility that the quantum vacuum energy density of the Casimir effect contributes to a (local) gravitational vacuum energy density.
Second, the possibility that
a change in the gravitational coupling implies a change in the cosmological constant.
We parametrize these two possibilities in a covariant framework and show
that the next generation of
Casimir experiments does have a surprisingly good chance of exploring this parameter space.
\end{abstract}
\maketitle

\tableofcontents

%%%%%%%%%%%%%%%%%%%%%%%%%%%%%%%%%
\section{Preamble}

This article combines several fields of physics, which are not commonly related.
This increases the risk of misunderstanding.
To minimize this risk we start with two questions of fundamental physics, which we 
ask in terms of two phenomenological hypotheses:
\begin{itemize}
    \item[$H_{Q\leftrightarrow\Lambda}$:]
    Is the quantum vacuum energy density linked to the cosmological energy density?
    \item[$H_{\Lambda \leftrightarrow G}$:]
    Does a change of the gravitational coupling $ G$ imply a change in the cosmological coupling $ \Lambda$ and vice versa?
\end{itemize}
The first question $H_{Q\leftrightarrow\Lambda}$ 
has two aspects.
On the theoretical side there it is at the  core of the cosmological constant problem, which will be introduced in sub-section \ref{ssec:VacEn}.
On the experimental side of $H_{Q\leftrightarrow\Lambda}$, we will make connection to the Casimir effect, which is introduced in subsection \ref{ssec_Casimir}.
The link between these two aspects of $H_{Q\leftrightarrow\Lambda}$ is established and parametrized in 
subsection \ref{ssec_alpha}
The question $H_{\Lambda\leftrightarrow G}$ can be parametrized in terms of scale-dependent gravitational couplings,
which is introduced in \ref{ssec_SDG}.

In the section \ref{sec_TowardsExp} we show how the interplay between the parametrized versions of $H_{Q\leftrightarrow\Lambda}$ and $H_{\Lambda\leftrightarrow G}$ can be  tested experimentally.

%%%%%%%%%%%%%%%%%%%
\section{Introduction}
%%%%%%%%%%%%%%%%

%%%%%%%%%%%%%%%%
\subsection{Vacuum energy and the cosmological constant}
\label{ssec:VacEn}

The cosmological constant was introduced by Einstein~\cite{Einstein:1917} to prevent gravitational instability in general relativity (GR). While the constant's history is complex, the acceleration of cosmic expansion~\cite{SupernovaCosmologyProject:1997zqe, SupernovaSearchTeam:1998fmf, SupernovaSearchTeam:1998bnz} provides us with a clear value~\cite{Planck:2018vyg} $\Lambda_0 %=8\pi \rho_0 \frac{G_0}{c^4}
= \SI{1.1e-52}{\metre^{-2}}$. 
We may interpret $\Lambda_0$ as an energy density contributing via the Newton coupling $G_0$ and the speed of light $c$ to the matter part of Einstein's field equations,
\be\label{eq_rhoL}
\rho_{\Lambda_0}=\frac{\Lambda_0 c^4}{8 \pi G_0}= \SI{5.35e-10}{\joule/\metre^3}.
\ee

While the interpretation of $\Lambda_0$ in GR is clear, its origin remains elusive. According to quantum field theory, the zero-point energy~\cite{Zeldovich:1968ehl} of all fields of the standard model and beyond as well as the Higgs phase transition~\cite{Martin:2012bt} should contribute to the measured value of $\rho_{\Lambda_0}$. 
Zero-point energies of each particle's quantum field provide an infinite contribution to
the quantum energy density $\rho_{Q,0}$. 
Those contributions can be rendered finite  $\rho_{Q,0}(\kappa)$ by introducing regulators like momentum cutoffs $\kappa$.
We can express the ratio between the cosmological energy density $\rho_{\Lambda_0}$ and the regularized quantum contributions
$\rho_{Q,0}(\kappa)$
as a dimensionless ratio
\be\label{eq_CCP}
\Upsilon_0\equiv
\frac{\rho_{\Lambda_0}}{\rho_{Q,0}(\kappa)}=
\frac{\Lambda_0 c^3 \hbar^3}{8 \pi G_0 \kappa^4_0}=
\begin{cases}
     10^{-121} & {\text{for}} \; \kappa_0 = c m_p  \\
     10^{-55} & {\text{for}} \; \kappa_0 = c m_{Z}
\end{cases},
\ee
where $m_Z$ and $m_p$ are the mass of the Z boson and the Planck mass, defining a cutoff choice at the weak and Planck scale, respectively~\cite{Sola:2013gha}.
According to our current understanding, $\Upsilon_0$ should be of order one (or exactly zero).
Instead, the values in \eqnref{eq_CCP} are small but non-vanishing, leading 
to a severe fine-tuning problem that was coined \emph{cosmological constant problem} (CCP)~\cite{Weinberg:1988cp,Sola:2013gha}.

Resolving the CCP will most likely require physics beyond 
the present SM and GR framework, which gave motivation for research on possible alternative descriptions of quantum gravity. The list of approaches  is long, ranging from standard techniques, such as the functional renormalization group~\cite{Percacci:2007sz,Codello:2008vh,Reuter:1996cp,Reuter:2001ag}, to Planck scale fluctuations~\cite{Cree:2018mcx,Wang:2020cvm}, or holographic interpretations \cite{Verlinde:2010hp,Verlinde:2016toy}.

New insight (in particular from experiments) concerning the quantum origin of $\Lambda_0$, is urgently needed.
In the following we argue that progress can not only be made by astronomical measurements but also by carefully tuned laboratory experiments.

%%%%%%%%%%%%%%%%%%%%%%%%%
\subsection{The Casimir effect}
\label{ssec_Casimir}

A setting, where quantum fluctuations are much better under control (both experimentally and theoretically), is the Casimir effect~\cite{Casimir:1948}, which remains the only known laboratory manifestation of the quantum vacuum causing a detectable interaction between macroscopic bodies. Here, spatial boundary conditions limit the mode spectrum of vacuum (and thermal) electromagnetic fluctuations. 
Like in the CCP, one needs to introduce
a regulator when calculating the otherwise divergent quantum vacuum energy 
$E_{Q}(\kappa)$
in a given volume subject to  boundary conditions.
Contrary to the CCP, the dependence on 
the regulator $\kappa$ can be circumvented
by subtracting the corresponding quantum vacuum energy
in this volume without boundary conditions $E_{Q,0}(\kappa)$
giving the renormalized Casimir energy
\be\label{eq_EC}
E_C= E_Q(\kappa)-E_{Q,0}(\kappa).
\ee
For the ideal case of two infinitely extended, perfectly conducting, parallel plates at distance $a$ and zero temperature, one obtains
\be
\sigma_{C,\text{id}}=
\frac{E_{C,\text{id}}(a)}{A}=-\frac{\hbar c \pi^2}{720 a^3}\,,\label{eq:casimir_energy_ideal}
\ee
for the energy per unit area $A$~\cite{Casimir:1948}. 
Real materials show dispersion and their response to electromagnetic fields typically falls off $\propto \omega_p/\omega^{-2}$ for high frequencies $\omega$ and a plasma frequency $\omega_p$, eventually leading to transparency in the far UV. Spectral dielectric properties, curved geometries, roughness, temperature, 
dynamical effects,
and further experimental details can also be
 considered~\cite{Lifshitz:1956,Bordag:2014,Corona-Ugalde:2015mtw} and lead to further corrections to~(\ref{eq:casimir_energy_ideal}). The penetration of electromagnetic modes and their energy into a surface can be described by the effective penetration depth $\delta_p\sim c/\omega_p$ of a material. For metals $\delta_p\sim 100$\,nm. 

While the theory of the Casimir energy is well understood, computation of the local energy density $\rho_C$ poses  problems~\cite{DeWitt:1975ys,Deutsch:1978sc}. The assumed step in the dielectric functions between the gap and the bounding surface leads to divergences $\propto z^{-4}$, depending on the distance $z$ from the (Dirichlet) boundary~\cite{Milton:2011iy,Bartolo:2014eoa}. One possible solution is the introduction of  soft~\cite{Milton:2011iy,Murray:2015tim,Griniasty:2017iix} or movable~\cite{Ford:1998he,Butera:2013hja,Armata:2014ksa,Armata:2017pqn} walls that smooth out the discontinuity, and thereby eliminate divergences~\cite{Zelnikov:2021ysf}. However, lacking a detailed first-principles approach, 
quantitative predictions strongly depend on the assumptions taken. The same is true for the region inside the walls~\cite{Shayit:2021kgn}. Independent of these details, however, $\rho_C$ can be expected to be large near the boundaries, for which it could potentially contribute to gravitational interactions~\cite{Fulling:2007xa}.
Up to now, experiments have tested configurations in which the Casimir force exceeds the gravitational interaction by several orders of magnitude. The upcoming \cannex{} setup~\cite{Sedmik:2021iaw}, however, will for the first time probe a regime, in which Casimir and gravitational interactions are of equal strength.
This opens up the exciting possibility of testing gravity in physical conditions with modified vacuum energy density.
%%%%%%%%%%%%%%%%
\subsection{Parametrizing question $H_{Q\leftrightarrow\Lambda}$}
\label{ssec_alpha}

Despite their suggestively analogous names, it is not clear whether the cosmological vacuum energy density $\rho_{\Lambda_0}$ and
the quantum energy density $\rho_C$ are related at all, i.e. if the CCP actually exists.
%The seriousness of our lack of theoretical understanding of this relation is exemplified by the CCP (\ref{eq_CCP}).
To test possible relations between the
two densities, we formulate the following working hypothesis:\\
\emph{The net cosmological energy density $\rho_\Lambda$
is a function of the bare cosmological energy density $\rho_{\Lambda_0}$ and a quantum contribution $\rho_C$}. For small $\rho_C$ this
relation can be linearized 
\bea\nonumber
\rho_\Lambda&=&\rho_\Lambda(\rho_0, \rho_C)\\
&\approx & \rho_{\Lambda_0} -\alpha \,\rho_C\,. \label{eq_rhoT}
%\\ \nonumber
%&\approx& \rho_0 \pm \alpha\frac{\hbar c \pi^2}{720 a^4}.
\eea
 The free parameter of this hypothesis is  $\alpha$.
Since $\rho_C$ arises from (electromagnetic) vacuum fluctuations, which should fully contribute to a total energy density, even values such $|\alpha| =1$, are reasonable, leading to the CCP. This is, for example, the case in well-known cosmological models~\cite{Leonhardt:2019epj}. 
In contrast, topological arguments of loop diagrams~\cite{Jaffe:2005vp} might suggest that $\alpha$ is strongly suppressed, or zero.
Thus, a laboratory test of the hypothesis 
\eqref{eq_rhoT} would give valuable information in this respect.
Note that $\rho_{\Lambda_0}$ might change on cosmological time-scales. However, for the time-scales involved in an experimental setup, such a dependence is negligible.

%%%%%%%%%%%%%%%%%%%%%%%%%%%
\subsection{Scale-dependent gravitational couplings: $H_{\Lambda\leftrightarrow G}$}
\label{ssec_SDG}

The gravitational couplings
$G(k)=G(\vec x)$ and $\Lambda(k)=\Lambda(\vec x)$ have to be generalized to SD quantities as well.
Such couplings can be dealt with in a theoretical framework,
known as SD gravity. \footnote{
Note that in the recent literature
there have been numerous discussions on 
the interplay between the Casimir force and (quantum) effects of gravitational interactions
\cite{Quach:2015qwa,Hu:2016lev,Santos:2016fpx,Inan:2017qdt,Alessio:2020lpk}.
These are not directly related to the 
SD scenario discussed in our paper}
%%%%%%%%%%%%%%%%
%
SD couplings are common
in the effective field theory approach to quantum gravity 
\cite{Eichhorn:2018yfc,Percacci:2007sz,Codello:2008vh,Reuter:1996cp,Reuter:2001ag}.
In these approaches, the functional form 
of $G(k)$ and $ \Lambda(k)$ is
determined by renormalization group equations, analogous to most other quantum 
field theories~\cite{Gell-Mann:1954yli,Wilson:1973jj}.
The usual known values of these couplings 
in a weak curvature expansion are
then associated with the asymptotic values 
\bea\label{eq_SD0}
\Lambda_0 = \lim_{k\rightarrow 0} \Lambda(k)\quad \text{and}\quad
G_0= \lim_{k\rightarrow 0} G(k).
\eea
Even though a uniquely accepted running of gravitational couplings is not
available, we can extract useful information from the SD of gravitational
couplings in the close vicinity of the deep infrared (IR). In this regime one can expect power-law running of the gravitational couplings~\cite{Wetterich:2017ixo}.
Thus, we can expand the SD gravitational couplings around 
$k_0=0$ \cite{Koch:2020baj,Laporte:2022wbu}
\begin{eqnarray}\label{eq_Gk}
G(k)&=&G_0(1+g(k))=G_{0}\left(1+C_{1}G_{0}k^2\right)+\mathcal{O}(k^4),\\
\Lambda(k)&=&\Lambda_0(1+\lambda(k))=\Lambda_{0}\left(1+C_{3}G_{0}k^2\right)+\mathcal{O}(k^4).\label{eq_Lk}
\end{eqnarray}
Here, $C_i $ 
parametrize the first effects of running couplings, when we depart from
the classical IR limit. 
This expansion is in the same spirit at the spirit of the effective field approach presented in~\cite{Bonanno:2020qfu}.
Even quite different approaches to quantum gravity
(e.g.~\cite{Lambiase:2022xde})
can, in principle, be mapped to \eqnsref{eq_Gk}, \ref{eq_Lk}), resulting in different values for these parameters. Depending on the theory
the $C_i$ can be small or large~\cite{Wetterich:2018qsl}.
Up to now, no experimental limits exist.
To extract quantitative predictions from SD couplings like \eqnsref{eq_Gk}, \ref{eq_Lk}), we need to set the RG scale $k$ in terms of physical parameters `$(\vec x,\; a, \; \dots)$' of a given observation (or experiment)
like energy, mass, or plate distance
$
k \rightarrow k(\vec x,\; a, \; \dots)$.
This crucial procedure is known as scale-setting.
There exists a large variety of scale-setting methods in conventional quantum field theory \cite{Stevenson:1981vj,Brodsky:1982gc,Grunberg:1982fw,Wu:2013ei}
and in quantum gravity 
\cite{Reuter:2003ca,Koch:2010nn,Domazet:2012tw,Koch:2014joa,Koch:2016uso,Contreras:2016mdt,Rincon:2017ypd,Rincon:2017ayr,Canales:2018tbn,Held:2021vwd}.
The applicability of these methods depends strongly
on the physical and experimental context.
Scale-setting also plays an important role in
the attempts to solve the cosmological constant problem~\cite{Mahajan:2006mw} (see also  \cite{Sahni:2002kh,Nojiri:2016mlb,Wetterich:2017ixo,Canales:2018tbn}).
Note, that in its ``Critique of the Asymptotic Safety Program'' Donoghue raises the question, whether SD couplings are a universal concept
(meaning that $C_i$ are universal constants)
or whether this running is ``only'' subjective for a particular observable,  in
which case the constants $C_i$ are only applicable for a specific observable~\cite{Donoghue:2019clr}.
Either way, universal or subjective, obtaining observable input for these parameters is 
a novelty.

Independent of which quantum gravity model gives rise to \eqnsref{eq_Gk}, \ref{eq_Lk}), or which scale-setting method is used,
after the scale-setting the SD couplings can be written as local quantities.
Thus, the Einstein field equations only remain consistent if they are
 generalized for SD couplings~\cite{Reuter:2003ca}.
These read
\be\label{eq_Gmn}
G_{\mu \nu}=8 \pi \frac{G(k)}{c^4} T_{\mu \nu}- \Lambda(k) g_{\mu \nu} -\Delta t_{\mu \nu}\,,
\ee
where 
\be
\Delta t_{\mu \nu}= G(k) 
\left(
g_{\mu \nu} \nabla^2-\nabla_\mu \nabla_\nu
\right)\frac{1}{G(k)}.
\ee

Note that already at this point it 
is possible to establish a relation between $H_{Q\leftrightarrow\Lambda}$ and $H_{\Lambda\leftrightarrow G}$.
The seemingly innocent relation \eqref{eq_rhoT} has a notable consequence. Since the Casimir energy density  is a function of external dimensionful quantities and local coordintes $\vec x$, the same must be true for the resulting $\rho_\Lambda$. This dependence on local and global scales $k\equiv k(\vec x)$ renders $\rho_\Lambda(k)=\rho_\Lambda(\vec x)$ a scale-dependent (SD) quantity.
Consequently, the definition in terms of the gravitational couplings~\eqref{eq_rhoL} has to be generalized to allow for SD
\be\label{eq_rhoLk}
\rho_\Lambda(k) \equiv \frac{c^4 \Lambda(k) }{8 \pi G(k)}\,.
\ee

%%%%%%%%%%%%%%%%%%%
\section{Results}
%%%%%%%%%%%%%%%%%%%
\label{sec_TowardsExp}

%%%%%%%%%%%%%%%%%%%
\subsection{Weak modifications of the Netwon potential}
\label{ssec_ModNewton}

Now, we explore the Weak Gravitational curvature and Weak SD limit (WG-WSD) of~\eqref{eq_Gmn}.
For this purpose, we isolate the Ricci curvature tensor on the left-hand side
\bea\label{eq_Rmn}
R^\mu_{\;\nu}&=&8 \pi \frac{G(k)}{c^4}\left(T^\mu_{\;\nu}-\frac{\delta^\mu_{\;\nu} T}{2} \right)
\\ \nonumber
&&
+\Lambda(K) \delta^\mu_{\;\nu} + G(k)\left(\frac{\delta_{\;\nu}^\mu\nabla^2}{2}+\nabla^\mu \nabla_\nu \right)\frac{1}{G(k)}.
\eea
The WG-WSD is achieved by 
an expansion in formally small 
deviations from the flat Minkowsi background.
The line element is
expanded with the parameter $\epsilon_{\Phi}$
and the SD gravitational coupling is expanded with the parameter $\epsilon_G$.
For both bookkeeping 
parameters ($\epsilon_\Phi$, $\epsilon_G$) we consider 
\bea\label{eq_ExpandWeak}
ds^2&=&-\left(1+2\epsilon_\Phi \Phi(r,\theta, \phi)\right) c^2 dt^2
+\left(1-2\epsilon_\Phi \Psi(r,\theta, \phi)\right) dr^2\nonumber \\ 
&&
+ \left(1+2\epsilon_\Phi \Xi(r,\theta, \phi)\right)
r^2 d\Omega^2 + {\mathcal{O}}(\epsilon_\Phi^2),\\
\nonumber
G(k)&=&\epsilon_\Phi \left(G_0+ \epsilon_G \Delta G(k)
+{\mathcal{O}}(\epsilon_\Phi^2)
\right),\\
\nonumber
\Lambda(k)&\rightarrow & \epsilon_\Phi \Lambda(k),
\eea
where $(\Phi,\,\Psi,\,\Xi)$ are small deformations (potentials) of the flat metric
and $\Delta G \ll G_0 $ is the SD correction to 
the gravitational coupling $G_0$.
Further, for non-relativistic matter at rest
with energy density $\rho_M=\rho_M(r,\theta, \phi)$,
the stress energy tensor in spherical coordinates is
\be
\left(T^\mu_{\;\nu}-\frac{1}{2}\delta^\mu_{\;\nu} T \right)=\frac{\rho_M }{2}{\text{diag}}(-1,1,1,1).
\ee
With this and the expansion (\ref{eq_ExpandWeak}), the time-time component
of \eqnref{eq_Rmn} reads
\be\label{eq_Pois1}
\vec \nabla^2 \Phi(r,\theta, \phi)
= \frac{4 \pi}{c^4} G_0 \rho_M+
\frac{\epsilon_G}{\epsilon_\Phi}\frac{\vec\nabla^2\Delta G}{2 G_0} - \Lambda+{\mathcal{O}}(\epsilon_\Phi, \, \epsilon_G),
\ee
where a global factor of $-\epsilon_\Phi$ has been canceled.
This expansion is valid and physically reasonable in a regime where
$\epsilon_\Phi > \epsilon_G >\epsilon_\Phi^2$. 
Note that we also expect contributions $\sim \Delta G(k) \rho_M$ to \eqnref{eq_Pois1}. Such contributions do exist, but they are 
much smaller than the leading contributions shown in \eqnref{eq_Pois1}.
Next, we drop the formal expansion parameters, keeping in mind the smallness of the $\Delta G$ contribution.
For all practical purposes in the context of Casimir experiments, the cosmological term can be neglected with respect to the other terms.
Thus, the IR correction to the gravitational coupling is given in terms 
of the local Casimir energy density $\Delta G(k)= \Delta G(\rho_C(\vec x))\equiv \Delta G(\vec x)$.
Now, the Poisson equation (\ref{eq_Pois1})
can be solved by the usual Green's function method
\bea\label{eq_Phi}
\Phi(\vec x)
&=&-\frac{G_0}{c^4} \int_{V_1}\!{\rm d}^3x'\, \frac{ \tilde \rho_M(x')}{|\vec x -\vec x'|}+{\mathcal{O}}(\epsilon)\\ \nonumber
&=&
-\frac{G_0}{c^4} \int_{V_1}  \frac{ \rho_M(\vec x')}{|\vec x -\vec x'|}\,{\rm d}^3x'+\frac{\Delta G(\vec x)}{2G_0}+{\mathcal{O}}(\epsilon),
\eea
where $V_1$ is the region of the gravitational source. Here,
 we defined the apparent gravitational energy density
\be\label{eq_rhotilde}
\tilde \rho_M=\rho_m + c^4 \frac{\vec \nabla^2 \Delta G}{8\pi G_0^2}.
\ee
When we insert the result (\ref{eq_Phi}) into the geodesic equation for a test particle with position $x^\mu$,
we find, for the spatial components, 
in the non-relativistic limit
\be\label{eq_d2x}
\frac{d^2 \vec x}{dt^2}=-c^2\vec \nabla \Phi+{\mathcal{O}}(\epsilon_\Phi).
\ee
In order to relate this acceleration to a force in Newton's second law, we have two mass densities at our disposal.
The first one is the `original' mass density $\rho_M/c^2$, corresponding to the mass we find in the absence of $\Delta G$: $M_2=\int_{V_2}\!{\rm d}^3x\,\rho_M(x)/c^2$. The second is the apparent gravitational mass density (\ref{eq_rhotilde}), which extends the first one by the influence of the local vacuum energy density, and thus is the one that has to be used when calculating the gravitational force caused by one object on another, $\vec F_{12}=-\vec F_{21}$. 
To leading order in $(\epsilon_G, \epsilon_\Phi)$, the force, sensed by an extended body with volume $V_2$
is given by
\bea\label{eq_F12}
\vec F_{G,12}&=&-  \int_{V_2} d^3x_2\;
\tilde \rho_M(\vec x_2)
\vec \nabla \Phi(\vec x_2)\\ \nonumber
&=& - \frac{G_0}{c^4}
\int_{V_2} d^3x_2
 \int_{V_1} d^3x_1 \frac{ \tilde \rho_M\left(\vec x_1\right)
 \tilde \rho_M\left(\vec x_2\right)
 \left(\vec x_2 - \vec x_1\right)
 }{|\vec x_2 -\vec x_1|^3}.
\eea
The crucial difference between \eqnref{eq_F12} and the usual expression for the gravitational force between extended bodies is the distinction between
$\tilde \rho_M$ and $\rho_M$, which only appears for  $\vec \nabla^2\Delta G\neq 0$.

%%%%%%%%%%%%%%%%%%%%%%%%
\subsection{Density scale-setting:  linking $H_{Q\leftrightarrow\Lambda}$ and $H_{\Lambda\leftrightarrow G}$}
\label{ssec_ScaleSet}

To relate an effective action to
real observables, one has to choose the RG scale in terms of variables that describe the system under consideration.
Since these variables, are in many  cases local. Thus, the scale-setting can imply the breaking of local symmetries, unless one takes particular care throughout the scale-setting procedure.
Below, we show how a scale-setting
based on the definition of density
and a manifestly covariant scale-setting, both leading
to the same type of expression.

Since we are interested in the leading corrections to the asymptotic limit \eqref{eq_SD0}, we insert \eqnsref{eq_Gk}) and \eqref{eq_Lk} into the definition \eqref{eq_rhoLk} and expand to first order in $k^2$
\be\label{eq_rhoLkSer}
\rho_\Lambda(k)=\rho_{\Lambda_0} - k^2 c^4 \frac{(C_1-C_3)}{8 \pi}\Lambda_0+ {\mathcal{O}}(k^4).
\ee
Here, the second term is the aforementioned correction to the asymptotic definition~\eqref{eq_rhoL}. 
By construction, \eqnref{eq_rhoLkSer} is equal 
to $\rho_\Lambda$ defined in \eqnref{eq_rhoT}.
Thus, by subtracting \eqnsref{eq_rhoT}) and \eqref{eq_rhoLkSer}, we obtain the unique scale-setting that is consistent with 
the working hypothesis in \eqnref{eq_rhoT},
\be \label{eq_k2}
k^2= \alpha \frac{8 \pi \rho_C }{c^4(C_1-C_3)\Lambda_0}.
\ee
Reinserting this into the IR expansion \eqref{eq_Gk}, we find that the gravitational
coupling inherits a weak dependence on the electromagnetic Casimir energy density 
\be\label{eq_Ga}
G(k) \approx G_0\left(1+C_{1}G_{0}\alpha \frac{8 \pi \rho_C }{c^4(C_1-C_3)\Lambda_0}\right). 
\ee
Comparing \eqnref{eq_Ga} with the weak SD expansion in \eqnref{eq_ExpandWeak}, we can identify
\be\label{eq_DG}
\vec \nabla^2\Delta G \approx 
\alpha G_0^2 \frac{8 \pi C_{1}\vec \nabla^2\rho_C }{c^4  (C_1-C_3)\Lambda_0}.
\ee
This correction has to be inserted into \eqnref{eq_Phi} when calculating the modified gravitational potential or the induced force between two objects according to \eqnref{eq_F12}.
The three phenomenological parameters in the following discussion will be $(\alpha,\, C_1,\,C_3)$.

The density induced scale-setting combines the expansions (\ref{eq_Gk}, \ref{eq_Lk}), which are consistent with covariant equations, with the relation (\ref{eq_rhoT}), which is explicitly not covariant.
This is OK as long as one understands
(\ref{eq_rhoT}) as a matter contribution to the equations of motion. 
There is, however, a way to
formulate the spirit of equation (\ref{eq_rhoT}) in a covariant language.
This will be discussed in the next subsection.

%%%%%%%%%%%%%%%%%%%%%%%%
\subsection{
Covariant scale-setting:
$H_{Q\leftrightarrow\Lambda}$ and $H_{\Lambda\leftrightarrow G}$ unified}
\label{ssec_ScaleSetCov}

%%%%%%%%%%%%%%%%%%%%%%%%
%\subsubsection{

%}

For this we give
the effective action of 
SD gravity coupled to matter.
In the low curvature expansion it can be written as
\be\label{eq_Gammak}
\Gamma_k=\int d^4x \sqrt{-g}\left(c^4
\frac{R-2\Lambda(k)}{16 \pi G(k)}+{\mathcal{L}}_m(\phi,k)\right).
\ee
Here, ${\mathcal{L}}_m(\phi,k)$ is the scale-dependent effective Lagrangian of Standard Model fields $\phi$. For the Casimir effect, the relevant field 
is the electromagnetic $U(1)$ gauge field.
In the spirit of the expansion (\ref{eq_Gk}, \ref{eq_Lk}) in the gravitational sector,
we can also expand the 
electromagnetic Lagrange density 
\ref{eq_Lk})
\be\label{eq_LmIR}
{\mathcal{L}}_m(\phi,k)={\mathcal{L}}_{m,0}(\phi)+
k^2 {\mathcal{L}}_{m,1}(\phi)+ \dots
\ee
Here, the densities
of the effective electromagnetic Lagrangian are
\bea\label{eq_Lm}
{\mathcal{L}}_{m,0}&=& \frac{(\vec E^2-\vec B^2)}{2}+ a (\vec E^2-\vec B^2)^2+ a^* (\vec E \cdot \vec B)^2 + \dots\\ \nonumber
{\mathcal{L}}_{m,1}&=& \alpha_1 \frac{(\vec E^2-\vec B^2)}{2}+
\alpha_2 a (\vec E^2-\vec B^2)^2+
\alpha_3 a^* (\vec E \cdot \vec B)^2 + \dots.
\eea
Here, we defined the usual (one-loop) prefactors 
\cite{Bretn2010}
%???cite
%https://iopscience.iop.org/article/10.1088/1742-6596/229/1/012006/pdf
\be
a=\frac{2\alpha_{EM}^2}{m_e^4}, \quad a*=\frac{7}{4}a.
\ee
Now, the expansions (\ref{eq_LmIR}, \ref{eq_Lm}) replace the hypothesis (\ref{eq_rhoT}) and the coefficients $\alpha_i$ take the role of the parameter $\alpha$.
To avoid confusion with the notation, remember that $\alpha_{EM}$ is the electromagnetic coupling constant.
With 

In terms of the action, the cosmological constant problem becomes the question ``how and to which extend quantum modes of 
${\mathcal{L}}_m(\phi,k)$  contribute to $\Lambda(k)$''.
This question is now parametrized in terms of $\alpha_i$.
Varying (\ref{eq_Gammak}) with respect to the metric field $g_{\mu \nu}$ gives rise to the field equations (\ref{eq_Gmn}).
It is not clear whether general covariance is broken by quantum gravity~\cite{Anber:2009qp}, but it is certainly a feature, we would like to perserve.
The general covariance of the system,
even after the scale-setting, can be assured by the variational scale-setting prescription~\cite{Koch:2014joa}, which complements (\ref{eq_Gmn}) with
the condition
\be\label{eq_setk}
\frac{\partial}{\partial k^2} \left(c^4
\frac{R-2\Lambda(k)}{16 \pi G(k)}+{\mathcal{L}}_m(\phi,k)\right)=0.
\ee
Inserting the IR expansions  (\ref{eq_Gk}, \ref{eq_Lk}, \ref{eq_LmIR}) into the covariant scale-setting condition, we can solve (\ref{eq_setk}) for the optimal scale
\bea\label{eq_k2opt}\nonumber
k^2_{opt}&=&\frac{{-16 \pi G_0 \mathcal{L}}_{m,1}(\phi)+c^4(C_1(R-2\Lambda_0)+ 2 C_3 \Lambda_0)
}{2 c^4 G_0 (C_1^2 (R-2\Lambda_0)-C_2 R
+ (C_2 + C_1 C_3
-C_4 )2\Lambda_0)
}\\
&\approx& 
\frac{1}{2 C_1}\left( 
\frac{1}{G_0}+\frac{8 \pi {\mathcal{L}}_{m,1}}{(C_1-C_3) c^4 \Lambda_0}
\right)
\eea
In the second line we have neglected the $\sim R$ and higher order $\sim (C_2, \, C_4)$ contributions.
Interestingly, the only local contribution to this optimal scale, comes from the matter Lagrangian ${\mathcal{L}}_{m,1}(\phi)$. 
For the purpose of the Casimir experiment, we are not interested in the dynamics of the electromagnetic field itself. Therefore, the electromagnetic Lagrangian enters the equations only in terms of a local background value. This
background value is obtained
from a summation over all modes consistent with the Casimir boundary conditions
\bea\nonumber
\langle{\mathcal{L}}_{m,1} \rangle_{bg} &=& 
\alpha_1 \left\langle \frac{(\vec E^2-\vec B^2)}{2}\right\rangle_{bg}+
\alpha_2 a \left\langle (\vec E^2-\vec B^2)^2\right\rangle_{bg} + \dots\\
&=& 
\alpha_1 \rho_C(x)+ {\mathcal{O}}\left(\rho_C^2\right).\label{eq_Lbg}
\eea
Here, we have used the fact,
that in the Casimir setting 
the magnetic modes do not contribute $\vec B^2=\vec B\cdot \vec E=0$, while
the summation over all electric modes gives
$\langle \vec E^2/2\rangle=\rho_C(x)$.
Note that using this summation, subject to the boundary conditions, breaks  Lorentz invariance.
This is, however, not a problem since it is a direct consequence of the boundary conditions of the experimental setup.
With this, the local background value of the optimal scale
is
\be\label{eq_koptav}
\left\langle 
k^2_{opt}
\right\rangle_{bg}=
k_0^2 +
\alpha_1
 \frac{ 4 \pi \rho_C(x) }{c^4 \Lambda_0 (C_1^2-C_1 C_3)}+{\mathcal{O}}\left(\rho_C^2\right).
\ee
In this relation, we have summarized all constant terms into~$k_0^2$. This constant $k_0^2$ will not contribute to the equations of motion.
The value (\ref{eq_koptav}) can then be used when solving the field equations (\ref{eq_Gmn}), or their non-relativistic approximation (\ref{eq_F12}). 
For (\ref{eq_rhotilde}) we need $\vec \nabla^2 \Delta G$.
With (\ref{eq_koptav})
and (\ref{eq_Gk}) we find
\bea\label{eq_DGCov}
\vec \nabla^2 \Delta G &=&C_1 G_0^2 \vec \nabla^2\langle k_{opt}^2 \rangle\\ \nonumber
&=&
\alpha_1 \frac{ 4 G_0^2 \pi }{c^4  \Lambda_0 (C_1- C_3)}\vec \nabla^2 \rho_C(x)
+{\mathcal{O}}(\rho_C^2)
\eea
Now, we can compare the result
of this the formal covariant scale-setting (\ref{eq_DGCov}) with the result of the density-driven scale-setting (\ref{eq_DG}).
It turns out that in the end, the local part of the covariant scale-setting corresponds exactly to the density scale-setting~(\ref{eq_k2}) if one relabels the proportionality constant  of our hypothesis
\be
 \alpha  = \frac{\alpha_1}{2 C_1}.
\ee
This is a remarkable result. It shows that by assuming scale-dependence of all couplings (\ref{eq_Gk}, \ref{eq_Lk}, and \ref{eq_LmIR}) in the effective action, combined
with the covariant scale-setting (\ref{eq_setk})
gives the same result as the combination of the hypotheses $H_{Q\leftrightarrow\Lambda}$ and $H_{\Lambda \leftrightarrow G}$. The hypotheses can thus be covariantly unified into a single hypothesis of a scale-dependent Lagrangian with covariant scale-setting (\ref{eq_setk}).
Thus the two hypotheses look independent, but they are essentially the same concept.

%Note that even though, the final result (\ref{eq_DG}) does not look explicitly covariant, there is no reason to worry: Remember, for example, that even though the metric solution of the interior of a star does not have all the symmetries of GR in free space, it is still the result of covariant field equations.
The result (\ref{eq_DG}) is the product of a covariant scale-setting procedure (\ref{eq_setk}) combined with covariant field equations (\ref{eq_Gmn}) which are then applied to the non-covariant boundary conditions of the Casimir problem.
Such a covariant method might also be applicable to different systems, e.g.
with  a mild time-evolution (running) of the cosmological vacuum energy, which has been explored in~\cite{Perez:2021ybe,SolaPeracaula:2022hpd}.\\

%%%%%%%%%%%%%%%%%%%
\subsection{First estimate of the experimental reach}

Now we discuss the perspective for experimental tests of the hypothesis (\ref{eq_rhoT}).
The three phenomenological parameters in this discussion are $\{\alpha,\, C_1,\,C_3\}$.
From the results in \eqnsref{eq_F12} and \eqref{eq_DG}) it is clear that any Casimir experiment, which is sensitive enough to also measure the gravitational attraction between the two plates~\cite{Sedmik:2021iaw}, will be suited to resolve, to some extend, the difference
between $\rho_M$ and $\tilde \rho_M$.
This will then allow to determine, or constrain $\alpha,\, C_1$, and $C_3$.
To obtain a glimpse on the experimental relevance of this effect, it is necessary to make further assumptions about experimental details.
For the force between the plates, we have to integrate in \eqnref{eq_F12} over the volume of the two plates.
Novel corrections to this force arise from non-vanishing $\rho_C(z)$ inside the plate material, while
the functional form of this energy density is irrelevant for this purpose.
When entering the plates, the
vacuum energy density is
assumed to drop exponentially 
from a starting value $\rho_{C,\text{id}}$ to zero 
\be\label{eq_Skin}
\rho_C(z)
= \rho_{C,\text{id}}\exp \left(- \frac{|z|}{\delta_p}\right).
\ee
Here, 
$\rho_{C,\text{id}}=\sigma_{C,\text{id}}/a$, 
is estimated by the average of the ideallized Casimir energy density~(\ref{eq:casimir_energy_ideal}).
Since \eqnref{eq_Skin} is a bold simplification of a still unknown, but likely more complicated functional form of $\rho_C(z)$~\cite{Bartolo:2014eoa,Murray:2015tim,Griniasty:2017iix,Shayit:2021kgn,Zelnikov:2021ysf},
we defer a detailed numerical integration of \eqnref{eq_F12}. Instead, we revisit the WG-WSD assumption of \eqnref{eq_ExpandWeak} to get a feeling of what we could expect to find for the phenomenological parameters. 
By inspecting the result (\ref{eq_F12}), it is reasonable to assume that the integrated energy density of the matter material $\rho_M$ is bigger than the correction due to the integral over SD of the gravitational coupling. Thus,
\be
1 \ll (8\pi G_0^2)\frac{ \int_{a/2}^{D}\! {\rm d}z \;\rho_M(z) }{
\int_{a/2}^{D}\!{\rm d}z\, c^2 |\vec \nabla^2 \Delta G|}
\approx \left|\frac{C_1-C_3}{2\alpha C_1^2}\right|\frac{ \delta_p \Lambda_0 \rho_M D}{\rho_{C,\text{id}}}.
\ee
Using \eqnsref{eq_DG}) and
\eqref{eq:casimir_energy_ideal}, and realistic exemplary values of ($z=a/2$, $a=10^{-5}~$m, $\rho_M/c^2=\rho_\text{gold}/c^2=19.3~{\rm g}/{\rm cm}^3/c^2$, $\delta_p=  10^{-7}~$m, $D=0.01~$m) in the above expression, we find that
\be\label{eq_boundEstimate}
\left|\alpha \frac{C_1^2}{C_1-C_3}\right|\ll 10^{-30}.
\ee
???ben is here
Note that $\delta_p$ and the boundary value $\rho_{C,\text{id}}$ refer to the assumed exponential attenuation model in \eqnref{eq_Skin}, which is subject to significant theoretical and experimental uncertainty.
Nevertheless, the skin depth was investigated quantitatively in experiments~\cite{Lisanti:2005}. Moreover, since our assumption of a constant $\rho_C$ between the plates and an exponential fall-off within the boundary underestimates the true volume where $\tilde{\rho}_M\neq0$ (i.e. where the SD modification of the gravitational force is sourced), the bound \eqref{eq_boundEstimate} is a \emph{worst case estimate}. 
%think that the error in this approach with respect to reality will amount to a mere few orders of magnitude, for which the bound \eqref{eq_boundEstimate} will remain approximately valid. 
This implies a very strong impact on the parameter space~$\{\alpha, C_1, C_3\}$.
A more realistic modeling of the experimental situation in this, and other configurations, such as the measurement of the potential $\Phi(z)$ [cf. \eqnref{eq_Phi}] with test particles, will be part of our future projects.

%%%%%%%%%%%%%%%%%%%
\section{Finally}
%%%%%%%%%%%%%%%%%%%

%%%%%%%%%%%%%%%%%%%%%%%
\subsection{Discussion}

We have shown how to obtain novel experimental insight into the possible connection between the quantum vacuum energy and the energy density corresponding to the cosmological coupling $\Lambda$.

The sensitivity in \eqnref{eq_boundEstimate} is, despite of its large uncertainty due to the simple assumed model for attenuation inside the material, overwhelmingly strong
when we compare them with standard quantum gravity corrections to the Newtonian potential~\cite{Hamber:1995cq}.
These leading corrections are typically
suppressed by the extremely small factors $r_S/a$ or $\lambda_p^2/a^2$, where $r_S$ and $\lambda_p$ are the Schwarzschild radius and the Planck length, respectively, and $a$ is a typical length scale.
Small corrections imply that a phenomenological pre-factor $\tilde \alpha$
(name chosen analogous to our $\alpha$)
of such corrections would be allowed to be very large
\be\label{eq_corrUsual}
 \frac{\sqrt{\hbar G_0/c^3}}{a}\approx  10^{-30} \quad \Rightarrow \quad \tilde \alpha \lessapprox 10^{+30}.
\ee
The reason for the discrepancy between
the usual expectation \eqnref{eq_corrUsual} and the strength of our result \eqnref{eq_boundEstimate} is fourfold:
\begin{itemize}
    \item[i)] Comparable order of $\rho_{\Lambda_0}$ and $\rho_C$.
    \item[ii)] The SD quantities $\rho_\Lambda \leftrightarrow \Lambda(k) \leftrightarrow G(k)$ are linked through the hypothesis $H_{Q\leftrightarrow\Lambda}$ and $H_{\Lambda\leftrightarrow G}$.
    \item[iii)]  $\Delta t_{\mu \nu}$
    contribution to the equations of motion.
    \item[iv)] $\Delta t_{\mu \nu}$ enhanced for small skin depth.
\end{itemize}
None of these four aspects are considered in the usual estimate~\cite{Hamber:1995cq}. Note that i--iv are not independent ad-hoc assumptions, but are rather natural consequences of the hypothesis~\eqref{eq_rhoT}.

We will interpret the result \eqref{eq_boundEstimate} in inverted order of the items above.
\begin{itemize}
    \item[iv)] \eqnref{eq_boundEstimate} has to be interpreted with care since~\eqnref{eq_Skin} has large theoretical uncertainties. \item[iii)] The result \eqref{eq_boundEstimate} would have to be recalculated if the $\Delta t_{\mu \nu}$ term was absent from the modified field equations~\eqref{eq_Gmn}, or if there would be additional non-minimal terms.
    \item[ii)] If $\Lambda(k)$ is only very weakly linked to $G(k)$ then $C_1\ll C_1-C_3$. Thus, \eqnref{eq_boundEstimate} could be satisfied by fine-tuning of the $C_i$. Quantum Gravity calculations do not support this possibility~\cite{Dona:2013qba,Dona:2014pla,Dona:2015tnf,Eichhorn:2017egq,Koch:2020baj}.
    \item[i)]
    Finally, except for the scenarios iv)-ii), there remains the possibility that \eqnref{eq_boundEstimate} provides an opportunity to experimentally test the relation between the quantum and cosmological energy densities, and thereby to possibly gain insight on the CCP~(\ref{eq_CCP}).
\end{itemize}
What can an experimental sensitivity of $\alpha \approx 10^{-30}$ teach us about the CCP (\ref{eq_CCP})? The CCP arises from the ambition to 
predict $\rho_\Lambda$ in terms of $\rho_Q$ such that $\rho_\Lambda=\rho_\Lambda(\rho_Q)$. 
Without loss of generality we can write this ambition in factorized form as
\be\label{eq_rhoLY}
\rho_\Lambda=\Upsilon(\rho_Q) \cdot \rho_Q,
\ee
%where we simply factored out a proportionality in $\rho_Q$.
%As $\Upsilon$ is a function of the effective local energy density $\rho_Q$, $\Upsilon$ itself cannot be a constant, even if in the absence of any Casimir energy ($\rho_\Lambda=\rho_\Lambda_0$), we obtain a vanishingly small value $\Upsilon(\rho_Q)=\Upsilon_0$.
In terms of \eqnref{eq_rhoLY} the CCP is the statement that $\Upsilon_0$ is an extremely small number (\ref{eq_CCP}).
The fact that $\Upsilon_0$ is a small number,
as measured in cosmology without additional Casimir energy contribution (i.e. $\rho_Q=\rho_{Q,0}$), 
does not imply that it is constant. It could be a function $\Upsilon=\Upsilon(\rho_Q)$. 
According to the definition (\ref{eq_EC}),
the quantum vacuum energy density $\rho_Q$, in turn, is for sure changing with additional small Casimir contributions
\be\label{eq_rhoQC}
\rho_Q=\rho_{Q,0}+ \rho_C\,,
\ee
and hence $\Upsilon_0=\Upsilon(\rho_{Q,0})$. 
We define the dependence
of the CCP on changes in the quantum energy density (\ref{eq_rhoQC}) in terms of the logarithmic derivative
\be\label{eq_Upsp}
\Upsilon'_0\equiv \left.\frac{d\Upsilon(\rho_Q)}{d \ln\left(\rho_Q\right)}\right|_{\rho_{C}=0}.
\ee
Inserting \eqnsref{eq_CCP}) and \eqref{eq_rhoT}
into \eqnref{eq_Upsp} allows us to relate
the observable $\alpha$ to $\Upsilon_0$ and $\Upsilon'_0$ via
\be\label{eq_alpha_CCP}
\alpha = - \Upsilon_0' - \, \Upsilon_0.
\ee
This relation clearly states that experimental insight on $\alpha$ coming from Casimir experiments can, without knowing $\Upsilon'_0$ from some theory, not give unambiguous information on the CCP. This may come as no surprise, as Casimir experiments probe a \emph{difference} in $\rho_Q$ according to \eqnref{eq_rhoQC}, while the CCP is caused by the \emph{absolute values} of $\rho_\Lambda$ (and $\rho_Q$). If such theoretical connection $\Upsilon'$ can be established, however, direct experimental investigations of the CCP by means of force metrology would be possible.

%%%%%%%%%%%%%%%%%%%%%%
\subsection{Clarifying comments}

\begin{itemize}
    \item {\bf The result isn't Lorentz invariant, is this contradictory or inconsistent?}\\
    One has to distinguish at which level the fundamental symmetries like Lorentz invariance are broken. While breaking this symmetry at the level of the equations of motion is typically considered conflictive, a breaking at the level of solution and its source terms is totally OK.
    Our results are derived from the Lorentz invariant system of equations (\ref{eq_Gmn}) and (\ref{eq_setk}). Breaking of Lorentz symmetry occurs only 
    at two points.
    First, when going to the Newtonian limit and second by considering a particular background configuration for the optimal scale (\ref{eq_Lbg}). \\
    {\it{``No, there is no inconsiteny with Lorentz symmetry, since the breaking only occurs at the choice of the source term of the physical system under consideration.''}}
    \item {\bf The result violates general covariance, is this contradictory or inconsistent?}\\
    The theory (\ref{eq_Gammak}) and its equations (\ref{eq_Gmn}) and (\ref{eq_setk}) are invariant under general coordinate transformations.\\ {\it{``Like explained in for the previous comment, a symmetry breaking at the level of (approximate) solutions is not in conflict with the consistency of the theory.}}
    \item {\bf Why should the two hypotheses $H_{Q\leftrightarrow\Lambda}$ and $H_{\Lambda\leftrightarrow G}$ be combined?}\\
    The two hypotheses seem like a combination of independent assumptions. The question arises, why should such assumptions be combined. However, as shown in subsection \ref{ssec_ScaleSetCov}, 
    $H_{Q\leftrightarrow\Lambda}$ and $H_{\Lambda\leftrightarrow G}$ have the same origin.\\
    {``\it{The two hypotheses are just to faces of a single concept, namely covariant scale-dependence with covariant scale-setting.}''}
    \item {\bf{ What if $\alpha_1=0$?}}\\
    It could happen, that the leading scale-dependence parameters of some of the expansions are actually exactly zero (e.g. $\alpha_1$). In this case, the expansions have to be continued to the subleading terms such as for example~$\alpha_2, \alpha_3$.\\
    {``\it{We have to consider subleading coefficients.}''}
\end{itemize}
%https://iopscience.iop.org/article/10.1088/1742-6596/229/1/012006/pdf

%%%%%%%%%%%%%%%%%%%%%%
\subsection{Conclusion}

We explored the 
hypothesis \eqref{eq_rhoT} that the cosmological energy density $\rho_\Lambda$ is influenced by changes in the quantum energy density in terms of the Casimir vacuum energy density $\rho_C$ ($H_{Q\leftrightarrow\Lambda}$).
The local nature of $\rho_C$ made it
then inevitable to introduce SD to the gravitational couplings.
This SD of the vacuum energy density in the gravitational sector, is then related to 
the frequently used link between the value of the gravitational coupling and the cosmological coupling,
which is our second hypothesis ($H_{\Lambda\leftrightarrow G}$).
In subsection \ref{ssec_ScaleSetCov} we showed, that the same result can be obtained from the concept of universal covariant SD.
SD, when minimally combined with diffeomorphism invariance, then led in the WG-WSD limit to a modification 
of the gravitational potential~\eqref{eq_Phi}.

This means, that experiments,
which are sensitive to both the gravitational force~\eqref{eq_F12} and the Casimir force, have the potential to actually test the hypotheses.
The scope of such tests is exemplified in the bound (\ref{eq_boundEstimate}).
Naturally, this inequality depends on the parameters of both hypothesis, namely  $\alpha$ for $H_{Q\leftrightarrow\Lambda}$ and $C_i$ for $H_{\Lambda\leftrightarrow G}$.
Eventually, our results may lead to new insight on the CCP~\eqref{eq_alpha_CCP}.
%
%%%%%%%%%%%%%%%%%%%%%%%%%
\section*{Acknowledgements}
We thank F. Intravaia for discussions on the intricacies of calculating $\rho_C$.
We further thank H.~Abele, C.~Gooding, C. Laporte, A.~Padilla, I.~A.~Reyes, and H.~Skarke for comments. This work was supported by the Austrian Science Fund (FWF): P 34240-N.

\bibliography{AllRefsCasimir,journaldefs}

\begin{thebibliography}{10}
\newcommand{\enquote}[1]{`#1'}
\providecommand{\url}[1]{\texttt{#1}}
\providecommand{\urlprefix}{URL }
\expandafter\ifx\csname urlstyle\endcsname\relax
  \providecommand{\doi}[1]{doi:\discretionary{}{}{}#1}\else
  \providecommand{\doi}{doi:\discretionary{}{}{}\begingroup
  \urlstyle{rm}\Url}\fi
\providecommand{\bibAnnoteFile}[1]{%
  \IfFileExists{#1}{\begin{quotation}\noindent\textsc{Key:} #1\\
  \textsc{Annotation:}\ \input{#1}\end{quotation}}{}}
\providecommand{\bibAnnote}[2]{%
  \begin{quotation}\noindent\textsc{Key:} #1\\
  \textsc{Annotation:}\ #2\end{quotation}}
\providecommand{\eprint}[2][]{\url{#2}}

\bibitem{Einstein:1917}
A.~Einstein.
\newblock \emph{Sitzungsberichte der K\"oniglich Preu\ss ischen Akademie der
  Wissenschaften (Berlin)} 142--152 (1917).
\bibAnnoteFile{Einstein:1917}

\bibitem{SupernovaCosmologyProject:1997zqe}
S.~Perlmutter \emph{et~al.} (Supernova Cosmology Project).
\newblock \emph{Nature} \textbf{391}(LBL-41172, LBNL-41172), 51 (1998).
\newblock \doi{10.1038/34124}.
\newblock \eprint{astro-ph/9712212}.
\bibAnnoteFile{SupernovaCosmologyProject:1997zqe}

\bibitem{SupernovaSearchTeam:1998fmf}
A.~G. Riess \emph{et~al.}
\newblock \emph{The Astronomical Journal} \textbf{116}(3), 1009 (1998).
\newblock \doi{10.1086/300499}.
\bibAnnoteFile{SupernovaSearchTeam:1998fmf}

\bibitem{SupernovaSearchTeam:1998bnz}
B.~P. Schmidt \emph{et~al.}
\newblock \emph{Astrophys. J.} \textbf{507}(1), 46 (1998).
\newblock \doi{10.1086/306308}.
\bibAnnoteFile{SupernovaSearchTeam:1998bnz}

\bibitem{Planck:2018vyg}
P.~Collaboration \emph{et~al.}
\newblock \emph{Astronomy \& Astrophysics} \textbf{641}, A6 (2020).
\newblock \doi{10.1051/0004-6361/201833910}.
\newblock \eprint{1807.06209}.
\bibAnnoteFile{Planck:2018vyg}

\bibitem{Zeldovich:1968ehl}
Y.~B. Zel'dovich, A.~Krasinski and Y.~B. Zeldovich.
\newblock \emph{Soviet Physics\textendash Uspekhi [translation of Uspekhi
  Fizicheskikh Nauk]} \textbf{11}, 381 (1968).
\newblock \doi{10.1007/s10714-008-0624-6}.
\bibAnnoteFile{Zeldovich:1968ehl}

\bibitem{Martin:2012bt}
J.~Martin.
\newblock \emph{Comptes Rendus Physique} \textbf{13}(6), 566 (2012).
\newblock \doi{10.1016/j.crhy.2012.04.008}.
\bibAnnoteFile{Martin:2012bt}

\bibitem{Sola:2013gha}
J.~Sol{\`a}.
\newblock \emph{Journal of Physics: Conference Series} \textbf{453}, 012015
  (2013).
\newblock \doi{10.1088/1742-6596/453/1/012015}.
\bibAnnoteFile{Sola:2013gha}

\bibitem{Weinberg:1988cp}
S.~Weinberg.
\newblock \emph{Rev. Mod. Phys.} \textbf{61}(1), 1 (1989).
\newblock \doi{10.1103/RevModPhys.61.1}.
\bibAnnoteFile{Weinberg:1988cp}

\bibitem{Percacci:2007sz}
R.~Percacci 111--128 (2007).
\newblock \eprint{0709.3851}.
\bibAnnoteFile{Percacci:2007sz}

\bibitem{Codello:2008vh}
A.~Codello, R.~Percacci and C.~Rahmede.
\newblock \emph{Annals Phys.} \textbf{324}, 414 (2009).
\newblock \doi{10.1016/j.aop.2008.08.008}.
\newblock \eprint{0805.2909}.
\bibAnnoteFile{Codello:2008vh}

\bibitem{Reuter:1996cp}
M.~Reuter.
\newblock \emph{Physical Review D: Particles and Fields}
  \textbf{57}(DESY-96-080), 971 (1998).
\newblock \doi{10.1103/PhysRevD.57.971}.
\newblock \eprint{hep-th/9605030}.
\bibAnnoteFile{Reuter:1996cp}

\bibitem{Reuter:2001ag}
M.~Reuter and F.~Saueressig.
\newblock \emph{Physical Review D: Particles and Fields}
  \textbf{65}(MZ-TH-01-27), 065016 (2002).
\newblock \doi{10.1103/PhysRevD.65.065016}.
\newblock \eprint{hep-th/0110054}.
\bibAnnoteFile{Reuter:2001ag}

\bibitem{Cree:2018mcx}
S.~S. Cree \emph{et~al.}
\newblock \emph{Physical Review D} \textbf{98}(6), 063506 (2018).
\newblock \doi{10.1103/PhysRevD.98.063506}.
\bibAnnoteFile{Cree:2018mcx}

\bibitem{Wang:2020cvm}
Q.~Wang and W.~G. Unruh.
\newblock \emph{Phys. Rev. Lett.} \textbf{125}(8), 089001 (2020).
\newblock \doi{10.1103/PhysRevLett.125.089001}.
\newblock \eprint{2008.09314}.
\bibAnnoteFile{Wang:2020cvm}

\bibitem{Verlinde:2010hp}
E.~P. Verlinde.
\newblock \emph{JHEP} \textbf{04}, 029 (2011).
\newblock \doi{10.1007/JHEP04(2011)029}.
\newblock \eprint{1001.0785}.
\bibAnnoteFile{Verlinde:2010hp}

\bibitem{Verlinde:2016toy}
E.~P. Verlinde.
\newblock \emph{SciPost Phys.} \textbf{2}(3), 016 (2017).
\newblock \doi{10.21468/SciPostPhys.2.3.016}.
\newblock \eprint{1611.02269}.
\bibAnnoteFile{Verlinde:2016toy}

\bibitem{Casimir:1948}
H.~B.~G. Casimir.
\newblock \emph{Proc Ned Ak Wet} \textbf{51}, 793 (1948).
\bibAnnoteFile{Casimir:1948}

\bibitem{Lifshitz:1956}
E.~Lifshitz.
\newblock \emph{J. Exp. Theor. Phys.} \textbf{2}(73), 334 (1956).
\bibAnnoteFile{Lifshitz:1956}

\bibitem{Bordag:2014}
M.~Bordag \emph{et~al.}
\newblock \emph{Advances in the {{Casimir Effect}}}.
\newblock {Oxford University Press} (2014).
\newblock ISBN 978-0-19-871998-4.
\bibAnnoteFile{Bordag:2014}

\bibitem{Corona-Ugalde:2015mtw}
P.~Corona-Ugalde \emph{et~al.}
\newblock \emph{Phys. Rev. A} \textbf{93}(1), 012519 (2016).
\newblock \doi{10.1103/PhysRevA.93.012519}.
\newblock \eprint{1511.07502}.
\bibAnnoteFile{Corona-Ugalde:2015mtw}

\bibitem{DeWitt:1975ys}
B.~S. DeWitt.
\newblock \emph{Phys. Rept.} \textbf{19}, 295 (1975).
\newblock \doi{10.1016/0370-1573(75)90051-4}.
\bibAnnoteFile{DeWitt:1975ys}

\bibitem{Deutsch:1978sc}
D.~Deutsch and P.~Candelas.
\newblock \emph{Phys. Rev. D} \textbf{20}, 3063 (1979).
\newblock \doi{10.1103/PhysRevD.20.3063}.
\bibAnnoteFile{Deutsch:1978sc}

\bibitem{Milton:2011iy}
K.~A. Milton.
\newblock \emph{Phys. Rev. D} \textbf{84}, 065028 (2011).
\newblock \doi{10.1103/PhysRevD.84.065028}.
\newblock \eprint{1107.4589}.
\bibAnnoteFile{Milton:2011iy}

\bibitem{Bartolo:2014eoa}
N.~Bartolo \emph{et~al.}
\newblock \emph{J. Phys. Condens. Matter} \textbf{27}(21), 214015 (2015).
\newblock \doi{10.1088/0953-8984/27/21/214015}.
\newblock \eprint{1410.1492}.
\bibAnnoteFile{Bartolo:2014eoa}

\bibitem{Murray:2015tim}
S.~W. Murray \emph{et~al.}
\newblock \emph{Phys. Rev. D} \textbf{93}(10), 105010 (2016).
\newblock \doi{10.1103/PhysRevD.93.105010}.
\newblock \eprint{1512.09121}.
\bibAnnoteFile{Murray:2015tim}

\bibitem{Griniasty:2017iix}
I.~Griniasty and U.~Leonhardt.
\newblock \emph{Phys. Rev. B} \textbf{96}(20), 205418 (2017).
\newblock \doi{10.1103/PhysRevB.96.205418}.
\newblock \eprint{1704.03078}.
\bibAnnoteFile{Griniasty:2017iix}

\bibitem{Ford:1998he}
L.~H. Ford and N.~F. Svaiter.
\newblock \emph{Phys. Rev. D} \textbf{58}, 065007 (1998).
\newblock \doi{10.1103/PhysRevD.58.065007}.
\newblock \eprint{quant-ph/9804056}.
\bibAnnoteFile{Ford:1998he}

\bibitem{Butera:2013hja}
S.~Butera and R.~Passante.
\newblock \emph{Phys. Rev. Lett.} \textbf{111}, 060403 (2013).
\newblock \doi{10.1103/PhysRevLett.111.060403}.
\newblock \eprint{1302.6139}.
\bibAnnoteFile{Butera:2013hja}

\bibitem{Armata:2014ksa}
F.~Armata and R.~Passante.
\newblock \emph{Phys. Rev. D} \textbf{91}(2), 025012 (2015).
\newblock \doi{10.1103/PhysRevD.91.025012}.
\newblock \eprint{1411.5347}.
\bibAnnoteFile{Armata:2014ksa}

\bibitem{Armata:2017pqn}
F.~Armata \emph{et~al.}
\newblock \emph{J. Phys. Conf. Ser.} \textbf{880}(1), 012064 (2017).
\newblock \doi{10.1088/1742-6596/880/1/012064}.
\newblock \eprint{1703.02875}.
\bibAnnoteFile{Armata:2017pqn}

\bibitem{Zelnikov:2021ysf}
A.~Zelnikov and R.~Krechetnikov.
\newblock \emph{Eur. Phys. J. Plus} \textbf{136}(7), 755 (2021).
\newblock \doi{10.1140/epjp/s13360-021-01714-3}.
\newblock \eprint{2103.16713}.
\bibAnnoteFile{Zelnikov:2021ysf}

\bibitem{Shayit:2021kgn}
A.~Shayit \emph{et~al.}
\newblock \emph{Int. J. Mod. Phys. A} \textbf{37}(19), 2241007 (2022).
\newblock \doi{10.1142/S0217751X2241007X}.
\newblock \eprint{2107.10439}.
\bibAnnoteFile{Shayit:2021kgn}

\bibitem{Fulling:2007xa}
S.~A. Fulling \emph{et~al.}
\newblock \emph{Phys. Rev. D} \textbf{76}, 025004 (2007).
\newblock \doi{10.1103/PhysRevD.76.025004}.
\newblock \eprint{hep-th/0702091}.
\bibAnnoteFile{Fulling:2007xa}

\bibitem{Sedmik:2021iaw}
R.~I.~P. Sedmik and M.~Pitschmann.
\newblock \emph{Universe} \textbf{7}(7), 234 (2021).
\newblock \doi{10.3390/universe7070234}.
\bibAnnoteFile{Sedmik:2021iaw}

\bibitem{Leonhardt:2019epj}
U.~Leonhardt.
\newblock \emph{Annals of Physics} \textbf{411}, 167973 (2019).
\newblock \doi{10.1016/j.aop.2019.167973}.
\bibAnnoteFile{Leonhardt:2019epj}

\bibitem{Jaffe:2005vp}
R.~L. Jaffe.
\newblock \emph{Physical Review D} \textbf{72}(2), 021301 (2005).
\newblock \doi{10.1103/PhysRevD.72.021301}.
\bibAnnoteFile{Jaffe:2005vp}

\bibitem{Note1}
Note that in the recent literature there have been numerous discussions on the
  interplay between the Casimir force and (quantum) effects of gravitational
  interactions \cite
  {Quach:2015qwa,Hu:2016lev,Santos:2016fpx,Inan:2017qdt,Alessio:2020lpk}. These
  are not directly related to the SD scenario discussed in our paper.
\bibAnnoteFile{Note1}

\bibitem{Eichhorn:2018yfc}
A.~Eichhorn.
\newblock \emph{Front. Astron. Space Sci.} \textbf{5}, 47 (2019).
\newblock \doi{10.3389/fspas.2018.00047}.
\newblock \eprint{1810.07615}.
\bibAnnoteFile{Eichhorn:2018yfc}

\bibitem{Gell-Mann:1954yli}
M.~{Gell-Mann} and F.~E. Low.
\newblock \emph{Physical Review} \textbf{95}, 1300 (1954).
\newblock \doi{10.1103/PhysRev.95.1300}.
\bibAnnoteFile{Gell-Mann:1954yli}

\bibitem{Wilson:1973jj}
K.~G. Wilson and J.~B. Kogut \textbf{12}, 75 (1974).
\newblock \doi{10.1016/0370-1573(74)90023-4}.
\bibAnnoteFile{Wilson:1973jj}

\bibitem{Wetterich:2017ixo}
C.~Wetterich.
\newblock \emph{Physics Letters B} \textbf{773}, 6 (2017).
\newblock \doi{10.1016/j.physletb.2017.08.002}.
\newblock \eprint{1704.08040}.
\bibAnnoteFile{Wetterich:2017ixo}

\bibitem{Koch:2020baj}
B.~Koch and C.~Laporte.
\newblock \emph{Physical Review D} \textbf{103}(4), 045011 (2021).
\newblock \doi{10.1103/PhysRevD.103.045011}.
\bibAnnoteFile{Koch:2020baj}

\bibitem{Laporte:2022wbu}
C.~Laporte \emph{et~al.}  (2022).
\newblock \eprint{2207.13436}.
\bibAnnoteFile{Laporte:2022wbu}

\bibitem{Bonanno:2020qfu}
A.~Bonanno, G.~Kofinas and V.~Zarikas.
\newblock \emph{Phys. Rev. D} \textbf{103}(10), 104025 (2021).
\newblock \doi{10.1103/PhysRevD.103.104025}.
\newblock \eprint{2012.05338}.
\bibAnnoteFile{Bonanno:2020qfu}

\bibitem{Lambiase:2022xde}
G.~Lambiase and F.~Scardigli.
\newblock \emph{Phys. Rev. D} \textbf{105}(12), 124054 (2022).
\newblock \doi{10.1103/PhysRevD.105.124054}.
\newblock \eprint{2204.07416}.
\bibAnnoteFile{Lambiase:2022xde}

\bibitem{Wetterich:2018qsl}
C.~Wetterich.
\newblock \emph{Physical Review D: Particles and Fields} \textbf{98}(2), 026028
  (2018).
\newblock \doi{10.1103/PhysRevD.98.026028}.
\newblock \eprint{1802.05947}.
\bibAnnoteFile{Wetterich:2018qsl}

\bibitem{Stevenson:1981vj}
P.~M. Stevenson.
\newblock \emph{Physical Review D: Particles and Fields}
  \textbf{23}(DOE-ER-00881-185), 2916 (1981).
\newblock \doi{10.1103/PhysRevD.23.2916}.
\bibAnnoteFile{Stevenson:1981vj}

\bibitem{Brodsky:1982gc}
S.~J. Brodsky, G.~P. Lepage and P.~B. Mackenzie.
\newblock \emph{Physical Review D: Particles and Fields}
  \textbf{28}(SLAC-PUB-3011, FERMILAB-PUB-83-040-T), 228 (1983).
\newblock \doi{10.1103/PhysRevD.28.228}.
\bibAnnoteFile{Brodsky:1982gc}

\bibitem{Grunberg:1982fw}
G.~Grunberg.
\newblock \emph{Physical Review D: Particles and Fields}
  \textbf{29}(Print-82-0721 (ECOLE POLY)), 2315 (1984).
\newblock \doi{10.1103/PhysRevD.29.2315}.
\bibAnnoteFile{Grunberg:1982fw}

\bibitem{Wu:2013ei}
X.-G. Wu, S.~J. Brodsky and M.~Mojaza.
\newblock \emph{Progress in Particle and Nuclear Physics, Vol 63, No 1}
  \textbf{72}(SLAC-PUB-15282.-CP3-ORIGINS-2013-001.-DIAS-2013-1), 44 (2013).
\newblock \doi{10.1016/j.ppnp.2013.06.001}.
\newblock \eprint{1302.0599}.
\bibAnnoteFile{Wu:2013ei}

\bibitem{Reuter:2003ca}
M.~Reuter and H.~Weyer.
\newblock \emph{Physical Review D: Particles and Fields}
  \textbf{69}(MZ-TH-03-20), 104022 (2004).
\newblock \doi{10.1103/PhysRevD.69.104022}.
\newblock \eprint{hep-th/0311196}.
\bibAnnoteFile{Reuter:2003ca}

\bibitem{Koch:2010nn}
B.~Koch and I.~Ramirez \textbf{28}, 055008 (2011).
\newblock \doi{10.1088/0264-9381/28/5/055008}.
\newblock \eprint{1010.2799}.
\bibAnnoteFile{Koch:2010nn}

\bibitem{Domazet:2012tw}
S.~Domazet and H.~Stefancic \textbf{29}, 235005 (2012).
\newblock \doi{10.1088/0264-9381/29/23/235005}.
\newblock \eprint{1204.1483}.
\bibAnnoteFile{Domazet:2012tw}

\bibitem{Koch:2014joa}
B.~Koch, P.~Rioseco and C.~Contreras.
\newblock \emph{Physical Review D: Particles and Fields} \textbf{91}(2), 025009
  (2015).
\newblock \doi{10.1103/PhysRevD.91.025009}.
\newblock \eprint{1409.4443}.
\bibAnnoteFile{Koch:2014joa}

\bibitem{Koch:2016uso}
B.~Koch, I.~A. Reyes and {\'A}.~Rinc{\'o}n \textbf{33}(22), 225010 (2016).
\newblock \doi{10.1088/0264-9381/33/22/225010}.
\newblock \eprint{1606.04123}.
\bibAnnoteFile{Koch:2016uso}

\bibitem{Contreras:2016mdt}
C.~Contreras, B.~Koch and P.~Rioseco.
\newblock \emph{Yamada Conference Lx On Research in High Magnetic Fields}
  \textbf{720}(1), 012020 (2016).
\newblock \doi{10.1088/1742-6596/720/1/012020}.
\bibAnnoteFile{Contreras:2016mdt}

\bibitem{Rincon:2017ypd}
{\'A}.~Rinc{\'o}n, B.~Koch and I.~Reyes.
\newblock \emph{Yamada Conference Lx On Research in High Magnetic Fields}
  \textbf{831}(1), 012007 (2017).
\newblock \doi{10.1088/1742-6596/831/1/012007}.
\newblock \eprint{1701.04531}.
\bibAnnoteFile{Rincon:2017ypd}

\bibitem{Rincon:2017ayr}
A.~Rincon and B.~Koch.
\newblock \emph{Yamada Conference Lx On Research in High Magnetic Fields}
  \textbf{1043}(1), 012015 (2018).
\newblock \doi{10.1088/1742-6596/1043/1/012015}.
\newblock \eprint{1705.02729}.
\bibAnnoteFile{Rincon:2017ayr}

\bibitem{Canales:2018tbn}
F.~Canales \emph{et~al.}
\newblock \emph{Journal of Cosmology and Astroparticle Physics}
  \textbf{2020}(01), 021 (2020).
\newblock \doi{10.1088/1475-7516/2020/01/021}.
\bibAnnoteFile{Canales:2018tbn}

\bibitem{Held:2021vwd}
A.~Held (Imperial/TP/2021/AH/04) (2021).
\newblock \eprint{2105.11458}.
\bibAnnoteFile{Held:2021vwd}

\bibitem{Mahajan:2006mw}
G.~Mahajan, S.~Sarkar and T.~Padmanabhan.
\newblock \emph{Physics Letters B} \textbf{641}(1), 6 (2006).
\newblock \doi{10.1016/j.physletb.2006.08.026}.
\bibAnnoteFile{Mahajan:2006mw}

\bibitem{Sahni:2002kh}
V.~Sahni \textbf{19}(IUCAA-08-2002), 3435 (2002).
\newblock \doi{10.1088/0264-9381/19/13/304}.
\newblock \eprint{astro-ph/0202076}.
\bibAnnoteFile{Sahni:2002kh}

\bibitem{Nojiri:2016mlb}
S.~Nojiri.
\newblock \emph{Modern Physics Letters A} \textbf{31}(37), 1650213 (2016).
\newblock \doi{10.1142/S0217732316502138}.
\newblock \eprint{1601.02203}.
\bibAnnoteFile{Nojiri:2016mlb}

\bibitem{Donoghue:2019clr}
J.~F. Donoghue.
\newblock \emph{Front. in Phys.} \textbf{8}, 56 (2020).
\newblock \doi{10.3389/fphy.2020.00056}.
\newblock \eprint{1911.02967}.
\bibAnnoteFile{Donoghue:2019clr}

\bibitem{Bretn2010}
N.~Bretón.
\newblock \emph{Journal of Physics: Conference Series} \textbf{229}(1), 012006
  (2010).
\newblock \doi{10.1088/1742-6596/229/1/012006}.
\newblock \urlprefix\url{https://dx.doi.org/10.1088/1742-6596/229/1/012006}.
\bibAnnoteFile{Bretn2010}

\bibitem{Anber:2009qp}
M.~M. Anber, U.~Aydemir and J.~F. Donoghue.
\newblock \emph{Phys. Rev. D} \textbf{81}, 084059 (2010).
\newblock \doi{10.1103/PhysRevD.81.084059}.
\newblock \eprint{0911.4123}.
\bibAnnoteFile{Anber:2009qp}

\bibitem{Perez:2021ybe}
J.~d.~C. Perez \emph{et~al.}
\newblock \emph{In} \emph{{16th Marcel Grossmann Meeting on~Recent Developments
  in Theoretical and Experimental General Relativity, Astrophysics and
  Relativistic Field Theories}} (2021).
\newblock \doi{10.1142/9789811269776_0137}.
\newblock \eprint{2110.07569}.
\bibAnnoteFile{Perez:2021ybe}

\bibitem{SolaPeracaula:2022hpd}
J.~Sola~Peracaula.
\newblock \emph{Phil. Trans. Roy. Soc. Lond. A} \textbf{380}, 20210182 (2022).
\newblock \doi{10.1098/rsta.2021.0182}.
\newblock \eprint{2203.13757}.
\bibAnnoteFile{SolaPeracaula:2022hpd}

\bibitem{Lisanti:2005}
M.~Lisanti, D.~Iannuzzi and F.~Capasso.
\newblock \emph{PNAS} \textbf{102}(34), 11989 (2005).
\newblock \doi{10.1073/pnas.0505614102}.
\bibAnnoteFile{Lisanti:2005}

\bibitem{Hamber:1995cq}
H.~W. Hamber and S.~Liu.
\newblock \emph{Phys. Lett. B} \textbf{357}, 51 (1995).
\newblock \doi{10.1016/0370-2693(95)00790-R}.
\newblock \eprint{hep-th/9505182}.
\bibAnnoteFile{Hamber:1995cq}

\bibitem{Dona:2013qba}
P.~Don{\`a}, A.~Eichhorn and R.~Percacci.
\newblock \emph{Physical Review D: Particles and Fields} \textbf{89}(8), 084035
  (2014).
\newblock \doi{10.1103/PhysRevD.89.084035}.
\newblock \eprint{1311.2898}.
\bibAnnoteFile{Dona:2013qba}

\bibitem{Dona:2014pla}
P.~Don{\`a}, A.~Eichhorn and R.~Percacci.
\newblock \emph{Canadian Journal of Physics} \textbf{93}(9), 988 (2015).
\newblock \doi{10.1139/cjp-2014-0574}.
\newblock \eprint{1410.4411}.
\bibAnnoteFile{Dona:2014pla}

\bibitem{Dona:2015tnf}
P.~Don{\`a} \emph{et~al.}
\newblock \emph{Physical Review D: Particles and Fields} \textbf{93}(4), 044049
  (2016).
\newblock \doi{10.1103/PhysRevD.93.129904}.
\newblock \eprint{1512.01589}.
\bibAnnoteFile{Dona:2015tnf}

\bibitem{Eichhorn:2017egq}
A.~Eichhorn.
\newblock \emph{Foundations of Physics} \textbf{48}(10), 1407 (2018).
\newblock \doi{10.1007/s10701-018-0196-6}.
\newblock \eprint{1709.03696}.
\bibAnnoteFile{Eichhorn:2017egq}

\bibitem{Quach:2015qwa}
J.~Q. Quach.
\newblock \emph{Physical Review Letters} \textbf{114}(8), 081104 (2015).
\newblock \doi{10.1103/PhysRevLett.114.081104}.
\newblock \eprint{1502.07429}.
\bibAnnoteFile{Quach:2015qwa}

\bibitem{Hu:2016lev}
J.~Hu and H.~Yu.
\newblock \emph{Physics Letters B} \textbf{767}, 16 (2017).
\newblock \doi{10.1016/j.physletb.2017.01.038}.
\newblock \eprint{1605.02193}.
\bibAnnoteFile{Hu:2016lev}

\bibitem{Santos:2016fpx}
A.~F. Santos and F.~C. Khanna.
\newblock \emph{International Journal of Theoretical Physics} \textbf{55}(12),
  5356 (2016).
\newblock \doi{10.1007/s10773-016-3156-y}.
\newblock \eprint{1605.08060}.
\bibAnnoteFile{Santos:2016fpx}

\bibitem{Inan:2017qdt}
N.~A. Inan, J.~J. Thompson and R.~Y. Chiao \textbf{65}(6-8), 1600066 (2017).
\newblock \doi{10.1002/prop.201600066}.
\bibAnnoteFile{Inan:2017qdt}

\bibitem{Alessio:2020lpk}
F.~Alessio, G.~Barnich and M.~Bonte.
\newblock \emph{JHEP} \textbf{02}, 216 (2021).
\newblock \doi{10.1007/JHEP02(2021)216}.
\newblock \eprint{2011.14432}.
\bibAnnoteFile{Alessio:2020lpk}

\end{thebibliography}
%%%%%%%%%%%%%%%%%%%%%%%%%%%%%%%%%%%%%%%%%%%
%%%%%%%%%%%%%%%%%%%%%%%%%%%%%%%%%%%%%%%%%%%

\end{document}